\documentclass[twocolumn,showpacs,preprintnumbers,amsmath,amssymb,prb, longbibliography]{revtex4-2}
\usepackage{graphicx}
\usepackage{dcolumn}
\usepackage{bm}
\usepackage{adjustbox}
\usepackage{tabularx}
\usepackage{booktabs}
\usepackage{calc} 
\usepackage{array}
\usepackage{hyperref}
\usepackage{ragged2e}
\usepackage{subcaption}
\usepackage{tikz}
\usetikzlibrary{arrows.meta, positioning}

\renewcommand{\j}{{\bf j}}

\def\lsim{\lower.35em\hbox{$\stackrel{\textstyle<}{\textstyle\sim}$}}
\def\gsim{\lower.35em\hbox{$\stackrel{\textstyle>}{\textstyle\sim}$}}

\begin{document}

\title{Prediction of the Hubbard $U$ parameter from scanning tunneling microscopy images of moiré systems using image recognition}
\author{Nachiket Tanksale$^1$ and Tobias Stauber$^{2}$$^,$$^{3}$ }
\affiliation{
$^{1}$ Universidad Internacional Men\'endez Pelayo, E-28040 Madrid, Spain\\
$^{2}$ Quantum Advanced Research Center (QuARC), CSIC, E-28049 Madrid, Spain\\
$^{3}$ Instituto de Ciencia de Materiales de Madrid (ICMM), CSIC, E-28049 Madrid, Spain
}
\date{\today}
\begin{abstract}
The atomistic Hubbard interaction $U$, representing the on-site Coulomb repulsion, serves as a pivotal parameter in theoretical models describing correlated systems, yet its precise experimental determination, especially in moiré systems, remains challenging. Scanning Tunneling Microscopy (STM) provides real-space images of the local density of states (LDOS), offering rich data sets that reflect the unique electronic structure of the material. Here, we introduce a systematic methodology for extracting the Hubbard U parameter directly from these LDOS images through the application of machine learning (ML) in the case of twisted bilayer graphene in the flat-band regime. Accurate regression of U is achieved despite the extremely high visual similarity between FT-LDOS images corresponding to different interaction strengths. Subsequent data analysis further suggests a gradual interaction-dependent redistribution of spectral weight, with a possible crossover scale of order $U/t \sim 1$. To assess robustness beyond idealized simulations, we introduce physically motivated perturbations that emulate experimental STM imperfections and demonstrate that augmentation-based training substantially improves generalization to noisy and symmetry-broken FT-LDOS images. 
\end{abstract}
\maketitle
\section{Introduction}
\label{sec:introduction}
The discovery of correlated insulating states and superconductivity in twisted bilayer graphene (TBG) has had a tremendous impact on the condensed matter community, revealing a phase diagram that closely resembles that of high-$T_c$ cuprate superconductors, including a strange-metal regime with linear-in-temperature resistivity near the superconducting dome.\cite{Cao18a,Yankowitz19,Moriyama19,Codecido19,Shen_2020,Lu19a,Chen19,Xu18,Guo18,Dodaro18,Baskaran18,Cao20} 
These phenomena are rooted in the emergence of nearly flat bands at magic twist angles, where structural complexity in moiré superlattices replaces chemical complexity as the main route to strong electronic correlations.\cite{Koshino18,Kang18,Isobe18,Zou18,Peltonen18,Kennes18,Guinea18,Gonzalez19,Thomson18,Carr18} 

During the last eight years, flat-band engineering has become a central theme in van der Waals heterostructures, as many two-dimensional crystals with low-energy excitations near high-symmetry points such as K, X, or M-point\cite{calugaru_moire_2025} can be tuned into correlated regimes by twisting, often summarized by the phrase ``moir\'e is more.''\cite{Peltonen18,Kennes18,Guinea18,Gonzalez19,Thomson18,Carr18} 
In TBG and related systems, strong Coulomb interactions drive cascade-like phase transitions at integer fillings, interaction-induced reconstruction of the band structure, and flavor (spin and valley) symmetry breaking, often accompanied by ferromagnetism at fractional band fillings.\cite{Codecido19,Chen19,Shen_2020,Lu19a,Yankowitz19,Moriyama19,Thomson18,Guinea18} 
At the same time, superconductivity appears extremely sensitive to screening and sample conditions, and both electron-phonon and purely electronic pairing mechanisms have been proposed.\cite{Liu18,You19,Zhang19,Gonzalez19,Gonzalez20,Cao20} 

Scanning tunneling microscopy (STM) and spectroscopy play a key role in disentangling these intertwined phases, as they provide direct, real-space access to the local density of states (LDOS) and correlated gaps in moiré systems at the atomic scale.\cite{Yankowitz19,Moriyama19,Codecido19,Shen_2020,Nuckolls23Nat,Kim23Nat} 
Recent STM experiments on magic-angle TBG have revealed Kekulé charge-density-wave order and shown that the STM signal, together with Chern-number information, can uniquely identify distinct correlated ground states.\cite{Nuckolls23Nat,Calugaru22PRL,Kwan_2024}  STM has thus proven to be an important technique to analyze moiré systems.\cite{Choi19,Xie2019,Mao2020,Choi21_band_flattening,Nuckolls23Nat,Lai2025} However, despite this wealth of spatially resolved data, quantitatively extracting microscopic interaction parameters, e.g., the effective on-site Hubbard repulsion $U$ directly from FT-LDOS images, remains a major open challenge and might open new opportunities also in view of the possibility to engineer magnetic phases in moiré systems.

Here, we want to explore whether the Hubbard interaction can, in principle, be inferred from experimentally accessible FT-LDOS data by fitting our model to the LDOS. To do this, we will use machine learning (ML), which has emerged as a powerful tool to tackle inverse problems in quantum many-body physics, where one aims to infer the underlying Hamiltonian or order parameters from measured observables.\cite{Carleo19RMP,Dawid_2025,Chertkov18PRX,Dubois22PRApp} 
Image-based techniques, especially convolutional neural networks (CNNs), have proven particularly effective for analyzing data-rich probes such as STM and photoemission, enabling automated detection of complex patterns and phase classification.\cite{Choudhary21SciData,Joucken22PRMat,Liu23PRB,RodriguezNieva19,Scheurer20,Sobral_2023,Taranto22} 
Moiré systems are especially well-suited for such approaches, since their enlarged unit cells expose intra-unit-cell physics, while their tunability allows high-dimensional datasets to be generated from a single sample.\cite{RodriguezNieva19,Scheurer20,Sobral_2023,Taranto22} 

Here, twisted bilayer graphene under hydrostatic pressure is modeled within a tight-binding description supplemented by a long-range Coulomb interaction and an on-site Hubbard term, and solved self-consistently at the Hartree-Fock level.\cite{sanchez_sanchez_nematic_2024,sanchez_sanchez_nonflat_2025,Moon13,Throckmorton12,Saito_2020,Stepanov_2020}  From the resulting eigenstates, we compute STM-accessible LDOS patterns and their Fourier transforms (FT-LDOS), which serve as "simulated STM-images" in momentum space for different values of $U$ in the flat-band regime. 
A deep CNN (both a custom architecture and a ResNet-18 variant) is then trained to regress the Hubbard $U$ directly from these FT-LDOS ``STM'' images, and interpretability tools and robustness analyses are employed to investigate the learned momentum-space representations and their stability under experimentally motivated perturbations.\cite{he2015deepresiduallearningimage,ILSVRC15,jacobgilpytorchcam} 

The paper is organized as follows. First, we introduce the tight-binding model and show how the FT-LDOS images are generated. Then, the training of the CNN is outlined. Last, we analyze the results with interpretability tools. We close with a summary and an outlook.  

\section{Tight-binding Model}
\label{sec:model}

We model twisted bilayer graphene under hydrostatic pressure, realizing a magic-angle-like condition at $\theta=3.5^\circ$, whose band structure closely resembles that at $\theta=1.16^\circ$ without pressure.\cite{sanchez_sanchez_nematic_2024} 
The non-interacting tight-binding Hamiltonian reads
\begin{align}
H_0 = -\sum_{n,m}\sum_{i,j}\sum_{\sigma}
  t_{n,m}^{i,j} \psi^\dagger_{n,i,\sigma} \psi_{m,j,\sigma} \;,
\end{align}
where $\psi^{(\dagger)}_{n,i,\sigma}$ are fermionic annihilation(creation) operators and the hopping parameters $t_{n,m}^{i,j}=t(|\mathbf{R}_n-\mathbf{R}_m+\mathbf{r}_i-\mathbf{r}_j|)$ only depend on the distance between lattice sites, with $\mathbf{R}_n$ the lattice vector of unit cell $n$ and $\mathbf{r}_i$ the position of site $i$ within the unit cell.\cite{Moon13}

We employ the Slater-Koster parametrization
\begin{align}
t(\mathbf{r}) &= V_{pp\pi} e^{\frac{a_0-r}{r_0}}
  \Bigg(1 - \left(\frac{\mathbf{r}\cdot\mathbf{e}_z}{r}\right)^2\Bigg) 
  + V_{pp\sigma} e^{\frac{d_\perp^0-r}{r_0}} 
  \left(\frac{\mathbf{r}\cdot\mathbf{e}_z}{r}\right)^2 \;,
  \label{eq:sk}
\end{align}
with carbon-carbon distance $a_0=1.42$\,\r{A}, decay length $r_0=0.319a_0$, equilibrium interlayer distance $d_\perp^0=3.34$\,\r{A}, and parameters $V_{pp\pi}=2.7$\,eV, $V_{pp\sigma}=-0.48$\,eV.\cite{Moon13} 
The compressed interlayer distance is $d_\perp=d_\perp^0/1.134$ for $\theta=3.48^\circ$.\cite{sanchez_sanchez_nematic_2024}

The full Hamiltonian is $H=H_0+H_\text{int}$ with interaction $H_\text{int}=H_V+H_U$ comprising long-range Coulomb ($H_V$) and on-site Hubbard ($H_U$) terms given by
\begin{align}
H_V &= \frac{1}{2} \sum_{n,m}\sum_{i,j}\sum_{\sigma,\sigma'}
  V_{n,m}^{i,j} \psi_{n,i,\sigma}^\dagger \psi_{m,j,\sigma'}^\dagger 
  \psi_{m,j,\sigma'} \psi_{n,i,\sigma} \;, \\
H_U &= \frac{U}{2} \sum_{n,i}\sum_{\sigma}
  \psi_{n,i,\sigma}^\dagger \psi_{n,i,\bar\sigma}^\dagger 
  \psi_{n,i,\bar\sigma} \psi_{n,i,\sigma} \;.
\end{align}
The Coulomb matrix elements are $V_{n,m}^{i,j}=v(|\mathbf{R}_n-\mathbf{R}_m+\mathbf{r}_i-\mathbf{r}_j|)$, implemented via the double-gated potential\cite{Throckmorton12}
\begin{align}
v(|\mathbf{r}|) &= \frac{e^2}{4\pi\epsilon_0\epsilon}
  \sum_n \frac{(-1)^n}{|\mathbf{r}+n\xi\mathbf{z}|} 
  \xrightarrow[r\gg\xi]{} \frac{e^2}{4\pi\epsilon_0\epsilon}
  \frac{2\sqrt{2}e^{-\pi r/\xi}}{\xi\sqrt{r/\xi}} \;.
  \label{eq:potential}
\end{align}
In the following, we will fix the screening length $\xi=10$\ nm and the dielectric constant $\epsilon=10$. The on-site Hubbard interaction $U$ can be thought of as a regularization of the Coulomb interaction at $\mathbf{r}=0$ and shall be varied from $0$ to $6$\,eV.\cite{Saito_2020,Stepanov_2020} 

\subsection{Hartree-Fock Solution}
We solve the interacting system within the restricted Hartree-Fock approximation, considering only spin-symmetric solutions using the FORGE package.\cite{sanchez_sanchez_nematic_2024,sanchez_sanchez_nonflat_2025,stauber_2025_17368106} The one-particle density matrix is thus restricted to solutions preserving all spatial, valley, and time-reversal symmetries to simplify neural-network training, and we suppress the spin index $\sigma$ from now on.\cite{Gonzalez19,Gonzalez20} In order to deal with a metallic state, we further choose an integer filling factor $\nu=3$ and set the Fermi energy $\epsilon_F$ to zero.

\begin{figure}[t]
    \centering
    \includegraphics[width=\columnwidth]{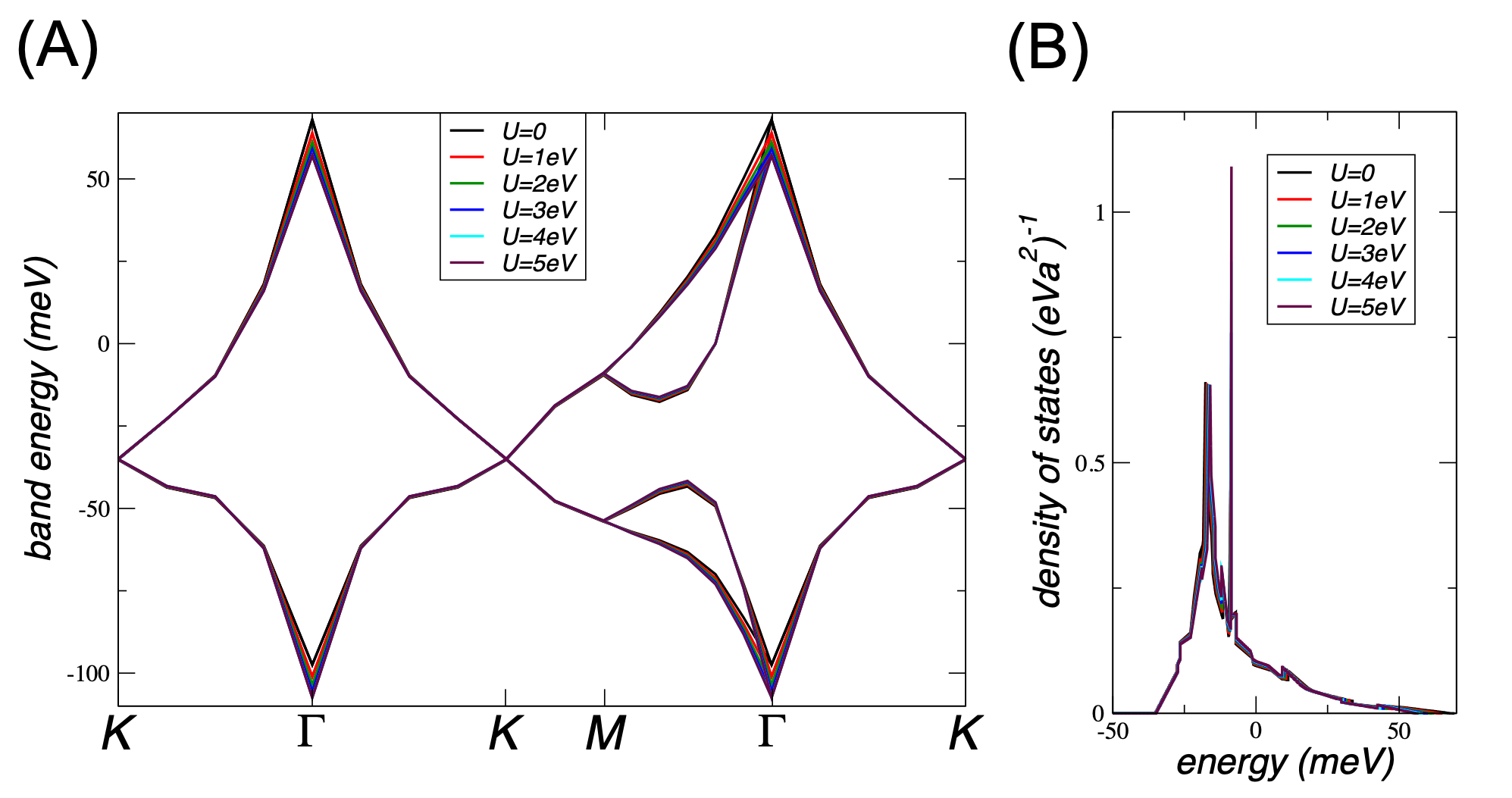}
    \caption{(A) Hartree-Fock bands at filling $\nu=3$ setting the Fermi energy $\epsilon_F$ to zero with $\epsilon=10$ obeying all symmetries, for $U=0$--$5$\,eV. (B) Density of states of the upper two flat bands for the same parameters.}
    \label{fig:bandsdos}
\end{figure}

Even though the atomistic Hubbard interaction is mainly associated with spin-related phenomena, there is a weak dependence on the band structure due to the localization of the wave-function around the AA-stacked region of the moiré unit cell.\cite{TramblydeLaissardiere2010} This is shown in Figure~\ref{fig:bandsdos}(a) where the self-consistent Hartree-Fock band structure is plotted for $U=0$--$5$\,eV. Only around the $\Gamma$-point, a slight variation can be seen. Consequently, the corresponding density of states of the upper two flat bands, defined by
\begin{align}
\rho_\nu(\epsilon) = \frac{\mathcal{N}}{A} \sum_{\mathbf{k},n}
  \delta(\epsilon-\epsilon_{\mathbf{k},n}^\nu) \;,
\end{align}
with $A$ the sample area, exhibits a similarly weak $U$-dependence [Fig.~\ref{fig:bandsdos}(b)]. Above, we also introduced the normalization $\mathcal{N}=\sqrt{3}\sin(\theta/2)/2$ ensuring $\int a^2\rho(\epsilon)\,d\epsilon$ counts bands within the energy window.

\subsection{Local Density of States}

The local density of states (LDOS) is directly proportional to the differential conductance measured by STM and therefore provides the natural bridge between our Hartree--Fock description and experimental images.\cite{Nuckolls23Nat,Kim23Nat,Calugaru22PRL} 
It is defined as
\begin{align}
\rho_\nu(\mathbf{r}_i,\epsilon)
  = \mathcal{N} \sum_{\mathbf{k},n}
    \left|\psi_{\mathbf{k},n}^\nu(\mathbf{r}_i)\right|^2
    \delta(\epsilon - \epsilon_{\mathbf{k},n}^\nu) \;,
\end{align}
where $\psi_{\mathbf{k},n}^\nu(\mathbf{r}_i)$ denotes the Hartree-Fock wave function at lattice site $\mathbf{r}_i$ with quantum numbers $(\mathbf{k},n)$ for filling factor $\nu$.

\begin{figure}[t]
    \centering
    \includegraphics[width=\columnwidth]{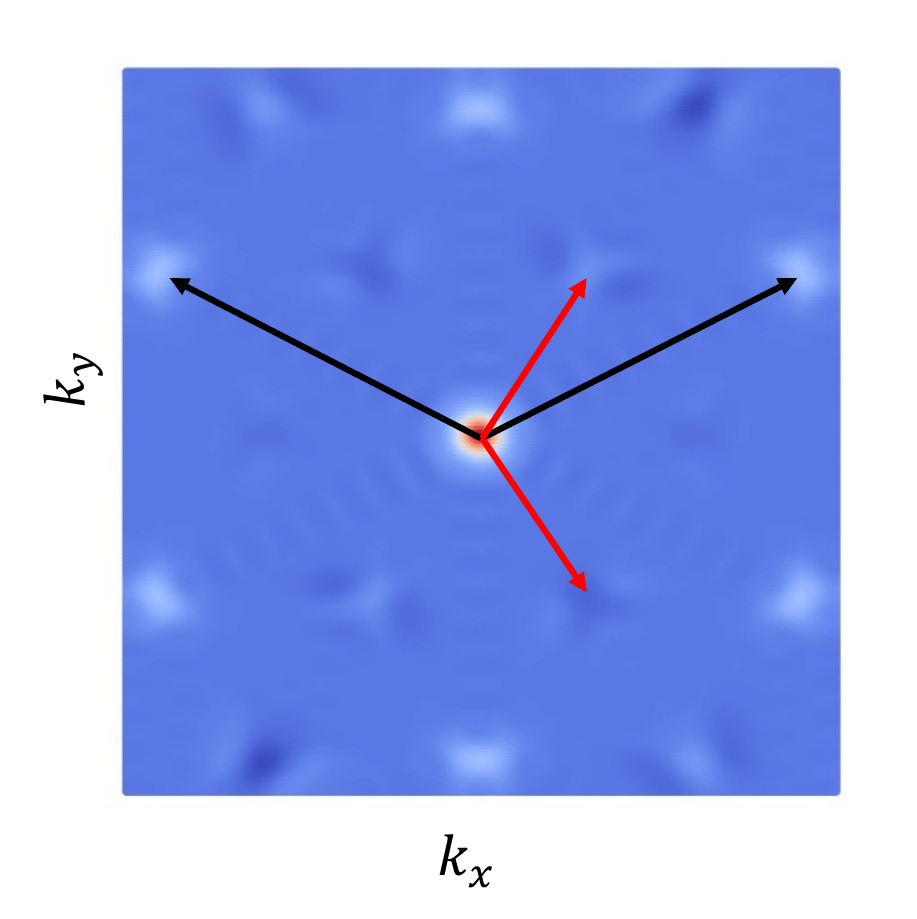}
    \caption{Fourier-transformed LDOS (FT-LDOS) for $\nu=-2$ and $\epsilon=12$  in the valley-coherence symmetry broken state, as discussed in Ref. \onlinecite{sanchez_sanchez_nematic_2024}. 
    The principal Bragg peaks (black arrows) are located at the reciprocal lattice vectors of the two rotated graphene layers, while the red arrows indicate the reciprocal vectors associated with the Kekulé superlattice composed of three graphene unit cells.}
    \label{DosBZ}
\end{figure}

Instead of working directly with $\rho_\nu(\mathbf{r}_i,\epsilon)$ in real space, we consider its discrete Fourier transform,
\begin{align}
\rho_\nu(\mathbf{k},\epsilon)
  = \sum_{\mathbf{r}_i} \rho_\nu(\mathbf{r}_i,\epsilon)
    e^{i\mathbf{k}\cdot \mathbf{r}_i} \;,
\end{align}
which corresponds to the momentum-space representation of STM conductance maps (FT-LDOS). The FT-LDOS representation enhances momentum-space features and Bragg-peak structures that evolve with interaction strength, making it particularly suitable for image-recognition approaches.
In a typical $\mathbf{k}$-space window, the FT-LDOS is characterized by six principal Bragg peaks defined by the two reciprocal lattice vectors $\mathbf{b}_{1/2} = \frac{4\pi}{3a}(\pm 3/2,\sqrt{3}/2)$, indicated by the black arrows in Fig. \ref{DosBZ}. More precisely, the six peaks arise from the superposition of the reciprocal lattice vectors of the two rotated graphene layers.

In symmetry-broken states with valley coherence and Kekulé order, additional Bragg peaks at
$\mathbf{b}_{1/2}^K = \frac{4\pi}{3a}(1/2,\pm\sqrt{3}/2)$
appear, reflecting the enlarged real-space unit cell.\cite{Nuckolls23Nat,Kwan_2024,sanchez_sanchez_nematic_2024,Sanchez26} 
This is indicated by the red arrows in Fig. \ref{DosBZ}, where the FT-LDOS of a valley-coherence symmetry broken state at $\nu=-2$ and $\epsilon=12$ is shown.\cite{sanchez_sanchez_nematic_2024}
Even though we focus our training on states preserving the full spatial and valley symmetry to simplify the ML task, the intensity and evolution of these Kekulé-related peaks in nearby parameter regimes may be crucial for understanding which FT-LDOS features are picked up by the convolutional neural networks.\cite{Taranto22,Sobral_2023}

\section{Methodology for image-recognition of FT-LDOS data}
\label{sec:methodology}

This section describes the generation of the theoretical dataset, the preprocessing of the Fourier-transformed local density of states (FT-LDOS), and the machine-learning architectures used to infer the interaction strength from simulated STM data.

\subsection{Dataset Generation from Hartree--Fock Theory}
\label{subsec:dataset_gen}

\begin{figure}[t]
    \centering
    \includegraphics[width=\columnwidth]{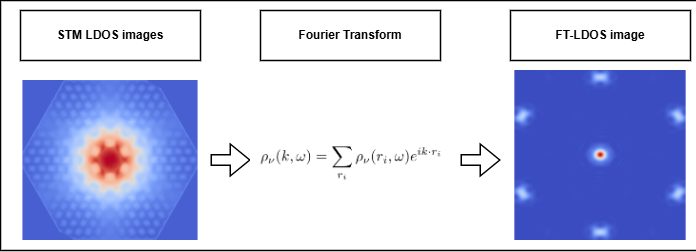}
    \caption{Local density of states (LDOS) for $\nu=3$ and $\epsilon=10$  and its Fourier-transformed local density of states (FT-LDOS).}
    \label{WorkFlow}
\end{figure}

We consider twisted bilayer graphene under hydrostatic pressure tuned such that a magic-angle-like flat-band regime is realized at $\theta \simeq 3.5^\circ$, with an effective band structure similar to that at $\theta \approx 1.16^\circ$ without pressure.\cite{sanchez_sanchez_nematic_2024} 
The moiré unit cell in this regime contains 1084 atoms (542 per layer), enabling self-consistent Hartree--Fock calculations including all bands and realistic long-ranged Coulomb interactions at high speed using the FORGE-package.\cite{stauber_2025_17368106}

For each value of the on-site Hubbard interaction $U$ in the range $0$-$6$\,eV, we compute eigenvalues and eigenstates at integer filling factor $\nu=3$ for fixed dielectric screening $\epsilon=10$.\cite{sanchez_sanchez_nematic_2024} 
We restrict ourselves to spin-symmetric solutions that preserve all spatial, valley, and time-reversal symmetries, avoiding additional symmetry breaking that would require separate neural-network training.
From these solutions, we generate STM LDOS maps and their FT-LDOS images at $\epsilon=\epsilon_F$, which serve as the input to the machine-learning models. This is illustrated in Fig. \ref{WorkFlow}.
To facilitate regression of the interaction strength, the training dataset is constructed using discrete anchor values of $U=0,1,2,3,4,$ and $5$ eV, yielding a total of $600$ FT-LDOS images. An additional holdout dataset of $66$ samples at fractional values of $U$ is generated to assess interpolation performance. This anchor-point strategy allows the models to learn the smooth evolution of FT-LDOS features across the interaction-strength manifold while providing an independent test of generalization within the trained parameter range.

\begin{figure}[t]
    \centering
    \includegraphics[width=\columnwidth]{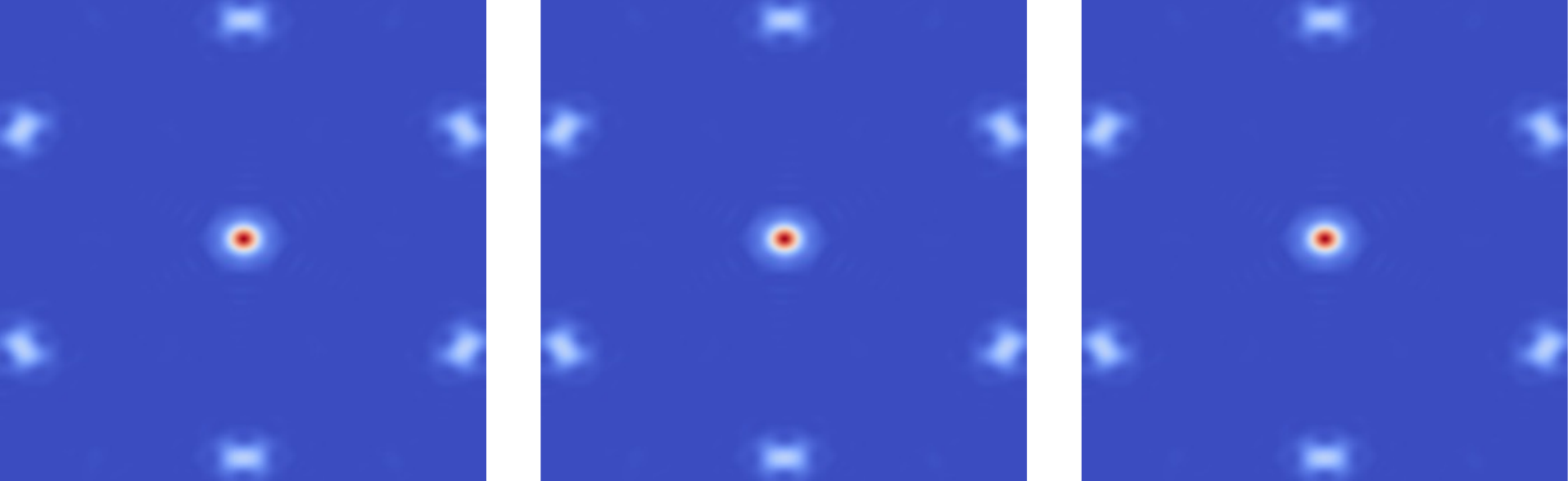}
    \caption{Fourier-transformed LDOS (FT-LDOS) for $\nu=3$ and $\epsilon=10$  in the symmetric state for their random values of the Hubbard interaction $U=0.59711$, $U=1.56952$, and $U=3.08776$. No difference is visible to the naked eye, which is also confirmed by cosine similarity, shown in Fig. \ref{fig:cosine_sim}.}
    \label{LDOS_3}
\end{figure}

\subsection{FT-LDOS Preprocessing}
\label{subsec:preprocessing}

The real-space LDOS maps are Fourier transformed to obtain $\rho_{\nu=3}(\mathbf{k},\epsilon=\epsilon_F)$, emphasizing reciprocal lattice vectors and interaction-sensitive momentum-space features commonly analyzed in STM experiments.\cite{Nuckolls23Nat,Calugaru22PRL}
The raw FT-LDOS images exhibit a dominant central moiré peak at $\mathbf{k}=0$ as well as pronounced Bragg peaks associated with the reciprocal lattice of the two (top and bottom) graphene lattices, see Fig.\ref{LDOS_3}.

To ensure numerical stability and efficient training while preserving physical content, each FT-LDOS image is rescaled to a resolution of $256\times256$ pixels and converted to grayscale, following standard practices in image-based learning of physical systems \cite{Choudhary21SciData,Joucken22PRMat}. The pixel intensities are then standardized to zero mean and unit variance using statistics computed exclusively on the training set, with the same transformation applied to the test and holdout datasets. This procedure prevents information leakage between training and evaluation sets and ensures consistent normalization across samples \cite{Goodfellow14}.

The full dataset is randomly partitioned into training and test subsets using an 80/20 split. Training and test splits were stratified with respect to the interaction parameter $U$ so that each discrete
$U$ value is proportionally represented in both sets. Although stratification is not strictly required for regression tasks with continuous targets, it is appropriate here due to the discrete sampling of $U$ in the generated dataset.

To assess robustness against sampling fluctuations, three independent splits with different random seeds are employed, and performance metrics are averaged where appropriate \cite{Carleo19RMP,Dawid_2025}.

\subsection{Neural-Network Architectures and Training Procedure}
\label{subsec:model_arch}

The inference task is formulated as a scalar regression problem, in which the goal is to predict the continuous interaction strength 
$U$ from a single FT-LDOS image. We employ two convolutional neural network (CNN) architectures to balance interpretability and representational power.

The first model is a custom-designed CNN consisting of three convolutional blocks, each comprising a two-dimensional convolution, batch normalization, and max pooling. All convolutions use $3\times3$
kernels with unit stride and zero padding, followed by rectified linear unit (ReLU) activations. The convolutional layers are followed by fully connected layers that map the extracted features to a single scalar output corresponding to the predicted interaction strength. This lightweight architecture is designed to preserve lattice symmetries in momentum space and to facilitate physical interpretation of learned features \cite{Goodfellow14}.

The second model is a modified ResNet-18 architecture pre-trained on the ImageNet dataset \cite{he2015deepresiduallearningimage,ILSVRC15}. The final classification layer is replaced by a fully connected regression head producing a single scalar output. The residual connections in ResNet-18 provide a robust feature extractor capable of capturing subtle variations in momentum-space textures, enabling comparison between task-specific and general-purpose architectures. The architectures of both models are depicted in Fig~\ref{fig:model_architectures}

\begin{figure*}[t]
\centering

\begin{subfigure}{0.48\textwidth}
\centering
\begin{tikzpicture}[
    node distance=0.55cm,
    block/.style={draw, rounded corners, align=center, minimum width=3.8cm, minimum height=0.75cm, font=\small},
    arrow/.style={-{Latex[length=2mm]}, thick}
]

\node[block] (input) {Input FT-LDOS image\\$1\times256\times256$};
\node[block, below=of input] (b1) {Conv block 1\\Conv(1,32) $\rightarrow$ ReLU $\rightarrow$ BN\\Conv(32,32) $\rightarrow$ ReLU $\rightarrow$ BN\\MaxPool};
\node[block, below=of b1] (b2) {Conv block 2\\Conv(32,64) $\rightarrow$ ReLU $\rightarrow$ BN\\Conv(64,64) $\rightarrow$ ReLU $\rightarrow$ BN\\MaxPool};
\node[block, below=of b2] (b3) {Conv block 3\\Conv(64,128) $\rightarrow$ ReLU $\rightarrow$ BN\\Conv(128,128) $\rightarrow$ ReLU $\rightarrow$ BN\\MaxPool};
\node[block, below=of b3] (pool) {Adaptive average pooling\\$128\times1\times1$};
\node[block, below=of pool] (flat) {Flatten};
\node[block, below=of flat] (norm) {$L_2$ feature normalization};
\node[block, below=of norm] (head) {Regression head\\Linear(128,256) $\rightarrow$ ReLU\\Dropout $\rightarrow$ Linear(256,1)};
\node[block, below=of head] (out) {Output\\$\hat{U}$};

\foreach \i/\j in {input/b1,b1/b2,b2/b3,b3/pool,pool/flat,flat/norm,norm/head,head/out}
    \draw[arrow] (\i) -- (\j);

\end{tikzpicture}
\caption{Custom CNN regressor.}
\label{fig:custom_cnn_arch}
\end{subfigure}
\hfill
\begin{subfigure}{0.48\textwidth}
\centering
\begin{tikzpicture}[
    node distance=0.65cm,
    block/.style={draw, rounded corners, align=center, minimum width=3.9cm, minimum height=0.75cm, font=\small},
    arrow/.style={-{Latex[length=2mm]}, thick}
]

\node[block] (input) {Input FT-LDOS image\\$1\times256\times256$};
\node[block, below=of input] (conv1) {ResNet18 stem\\Conv7$\times$7 $\rightarrow$ BN $\rightarrow$ ReLU\\MaxPool};
\node[block, below=of conv1] (l1) {Residual layer 1\\64 channels};
\node[block, below=of l1] (l2) {Residual layer 2\\128 channels};
\node[block, below=of l2] (l3) {Residual layer 3\\256 channels};
\node[block, below=of l3] (l4) {Residual layer 4\\512 channels};
\node[block, below=of l4] (pool) {Global average pooling};
\node[block, below=of pool] (flat) {Flatten\\512-dimensional feature vector};
\node[block, below=of flat] (norm) {$L_2$ feature normalization};
\node[block, below=of norm] (head) {Regression head\\Linear(512,1)};
\node[block, below=of head] (out) {Output\\$\hat{U}$};

\foreach \i/\j in {input/conv1,conv1/l1,l1/l2,l2/l3,l3/l4,l4/pool,pool/flat,flat/norm,norm/head,head/out}
    \draw[arrow] (\i) -- (\j);

\end{tikzpicture}
\caption{ResNet18-based regressor.}
\label{fig:resnet_arch}
\end{subfigure}

\caption{
Block diagrams of the two convolutional regression models used for Hubbard-$U$ prediction from FT-LDOS images.
Both models map a normalized FT-LDOS image to a scalar estimate $\hat{U}$.
}
\label{fig:model_architectures}
\end{figure*}

Both models are trained by minimizing the mean squared error loss between the predicted and true interaction strengths using the Adam optimizer. Model performance is evaluated using standard regression metrics, including the mean absolute error (MAE), root-mean-square error (RMSE), and coefficient of determination $R^2$. The hyperparameters used to train both models are listed in an Appendix~\ref{app:hyperparameters}.

\subsection{Model Interpretability and Auxiliary Analyses}
\label{subsec: model_interp}

To verify that the neural networks base their predictions on physically meaningful FT-LDOS features rather than numerical artifacts, we apply interpretability techniques adapted to regression tasks. Gradient-weighted class activation mapping (Grad-CAM) is used to identify regions in the final convolutional layers whose activations contribute most strongly to the predicted interaction strength \cite{jacobgilpytorchcam}. The resulting heatmaps highlight momentum-space regions that control the regression.

Guided backpropagation is employed to obtain high-resolution saliency maps at the input level, revealing which specific pixels in the FT-LDOS images influence the output most strongly \cite{jacobgilpytorchcam,Goodfellow14}. Combining Grad-CAM and guided backpropagation allows simultaneous identification of relevant momentum-space regions and fine-scale spectral features.

In addition to neural-network-based analysis, we perform auxiliary dimensionality reduction and baseline regression studies. Principal component analysis (PCA) is applied to the standardized FT-LDOS images to quantify the dataset's intrinsic dimensionality, and a ridge regression model trained on the leading principal components provides a linear baseline for comparison with the CNN results \cite{Carleo19RMP,Chertkov18PRX,Dubois22PRApp}.

\subsection{Data Augmentation and Robustness Evaluation}
\label{subsec: robust_eval}
The original FT-LDOS dataset is generated from a self-consistent Hartree--Fock calculation and therefore represents an idealized theoretical STM signal. Experimental STM measurements, however, typically contain instrumental noise, finite tip resolution, calibration uncertainties, scan-line artifacts, and weak symmetry-breaking distortions. To assess the robustness of the machine-learning models under more realistic conditions, we generated three additional augmented datasets from the original FT-LDOS images.

\begin{figure*}[t]
    \centering
    \begin{subfigure} [b]{0.24\textwidth}
        \includegraphics[width=\textwidth]{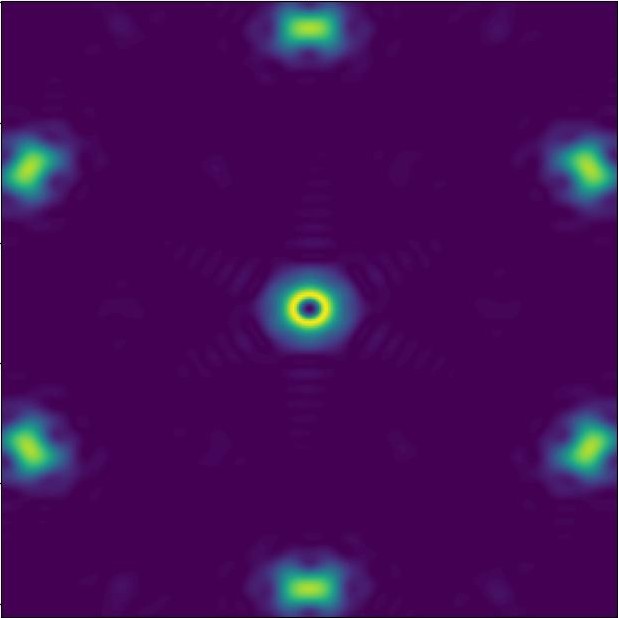}
        \caption{}
    \end{subfigure}
    \begin{subfigure} [b]{0.24\textwidth}
        \includegraphics[width=\textwidth]{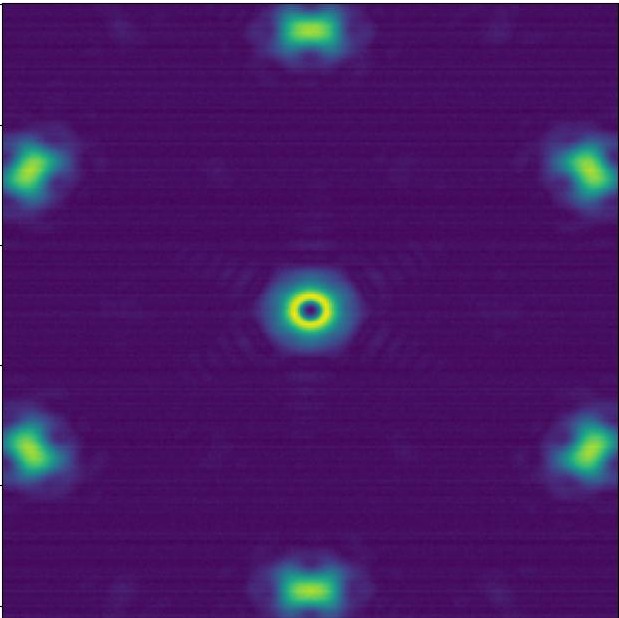}
        \caption{}
    \end{subfigure}
    \begin{subfigure} [b]{0.24\textwidth}
        \includegraphics[width=\textwidth]{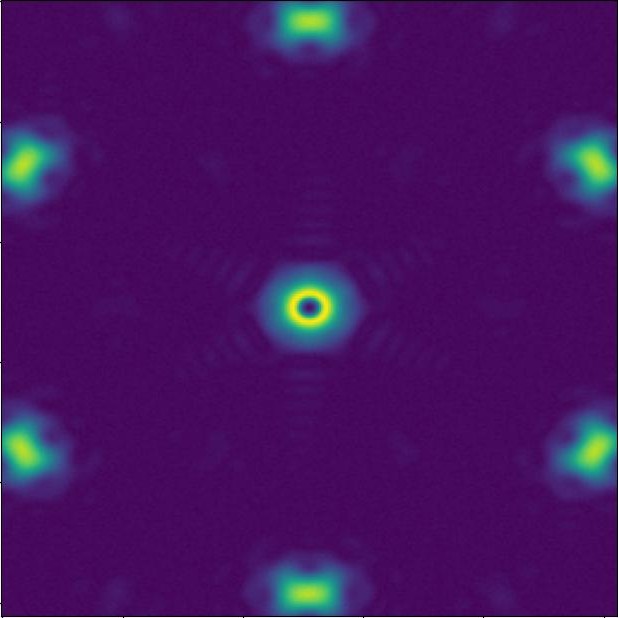}
        \caption{}
    \end{subfigure}
    \begin{subfigure} [b]{0.24\textwidth}
        \includegraphics[width=\textwidth]{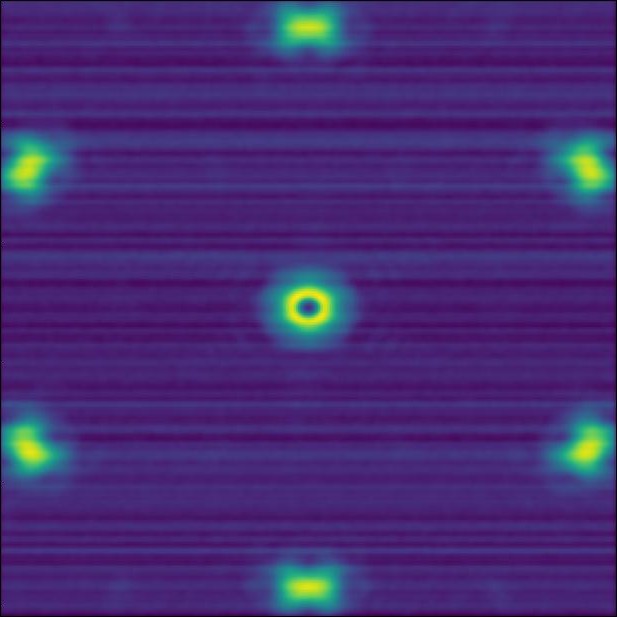}
        \caption{}
    \end{subfigure}
    \caption{
    Representative FT-LDOS images used in the robustness study. From left to right: (a) clean simulated image, (b) mildly perturbed image, (c) symmetry-breaking image, and (d) noisy image. The perturbations are designed to mimic realistic STM imperfections while preserving the dominant physical features of the FT-LDOS.
    }
    \label{fig:augmentation_examples}
\end{figure*}

Figure~\ref{fig:augmentation_examples} shows representative examples of the applied perturbations.

\begin{enumerate}
\item \textbf{Mild perturbation dataset.}
Weak physically motivated perturbations were applied while preserving the dominant FT-LDOS features. These include low-amplitude Gaussian noise, small intensity rescaling (gain variations), weak anisotropic blurring associated with non-ideal STM tip shapes, and correlated scan-line artifacts that mimic typical STM acquisition imperfections.
The perturbation amplitudes were intentionally restricted to regimes where the dominant FT-LDOS peaks remain visually identifiable, ensuring that the underlying physical label remains meaningful.

\item \textbf{Symmetry-breaking dataset.}
To investigate the sensitivity of the models to deviations from the ideal $C_3$ symmetry of the simulated images, small affine distortions were introduced. These perturbations emulate weak heterostrain, calibration errors, or other experimental effects that slightly deform the reciprocal-space structure without destroying the underlying physical features.
To emulate weak deviations from the ideal rotational symmetry of the simulated FT-LDOS images, we apply a small affine distortion

\begin{equation}
\mathbf{r}' =
\begin{pmatrix}
1+s & s/2\\
s/2 & 1-s
\end{pmatrix}
\mathbf{r},
\end{equation}

where $s$ is a small strain parameter randomly sampled from the interval
$s \in [0,0.015]$.
This transformation produces a weak $C_3$-symmetry-breaking distortion while preserving the dominant momentum-space features.

\item \textbf{Noisy dataset.}
A stronger perturbation regime was generated by combining scan-line artifacts, anisotropic blurring, gain variations, and stochastic noise. This dataset serves as a stress test of model robustness under more severe distribution shifts.
The perturbation amplitudes were intentionally restricted to regimes where the dominant FT-LDOS peaks remain visually identifiable, ensuring that the underlying physical label remains meaningful.
\end{enumerate}

Perturbation amplitudes were restricted to regimes where the dominant FT-LDOS peaks remain visually identifiable, ensuring that the underlying physical label remains meaningful.

The robustness evaluation was performed by training the machine-learning models on one dataset and evaluating them on another. In addition to the original clean dataset, cross-domain experiments were performed between the clean, mild, symmetry-breaking, and noisy datasets. This protocol provides a quantitative measure of model sensitivity to experimental imperfections and allows us
to assess whether the learned representation depends on idealized symmetries or remains stable under realistic perturbations.

A separate holdout dataset was used to evaluate interpolation capabilities beyond the discrete interaction values employed during training. Importantly, the holdout set contains interaction strengths not present in the training data. Successful prediction, therefore, requires interpolation in parameter space rather than memorization of the discrete training labels.

\section{Results and Discussion}
\label{sec:results}
Our main result is that the trained CNNs achieve highly accurate regression of the Hubbard interaction from simulated STM FT-LDOS images. In the following, we will discuss several aspects of our approach. 

\subsection{Evolution of FT-LDOS with Interaction Strength}
\label{subsec:evol_int_strength}

Before discussing machine-learning performance, it is instructive to examine how the Fourier-transformed LDOS (FT-LDOS) evolves as a function of the on-site interaction $U$. Across the range $U = 0-5$ eV, the dominant features of the FT-LDOS, namely the positions of the principal moiré Bragg peaks, remain unchanged, reflecting the robustness of the underlying moiré lattice geometry.

\begin{figure}[t]
    \centering
    \includegraphics[width=1.1\columnwidth, height=0.33\paperheight]{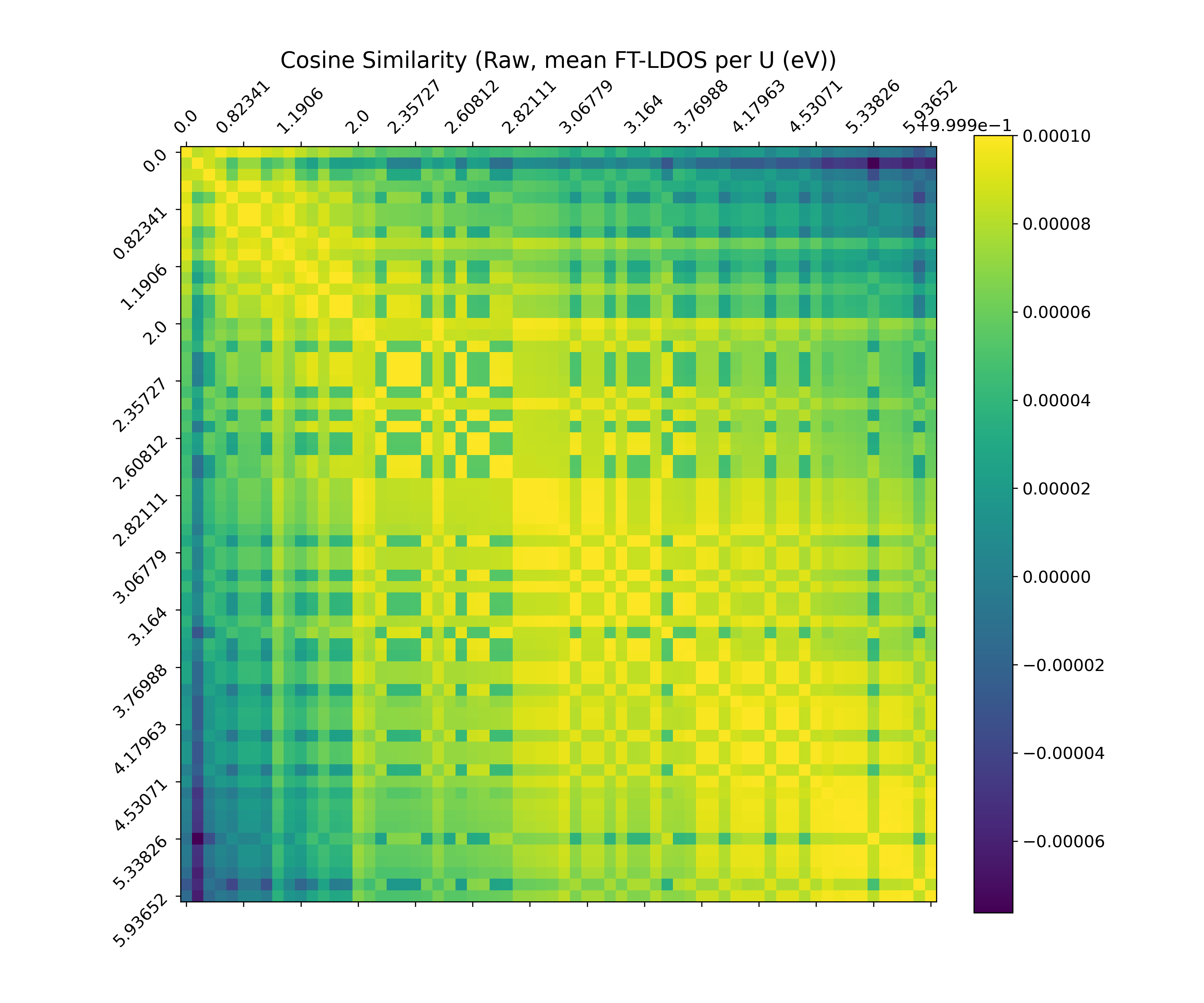}
    \caption{\textbf{Spectral Similarity:} Cosine similarity between the FT-LDOS magnitude maps across the surveyed range of $U$ (exceeding $99.98\%$) }
    \label{fig:cosine_sim}
\end{figure}

Fig. \ref{fig:cosine_sim} shows the pairwise cosine similarity between the mean FT-LDOS images corresponding to different interaction strengths. All values lie in the range 0.9998-1.0000, indicating that the raw images are extremely similar in a global linear sense. This high similarity is expected: varying $U$ primarily induces subtle redistributions of spectral weight rather than large structural changes in the Fourier-space LDOS. Consequently, simple distance metrics are insufficient to directly distinguish different $U$ values. The success of the neural-network models, therefore, demonstrates their ability to extract weak but systematic features from high-dimensional data that are not captured by naive similarity measures.

\begin{figure*}[t]
    \centering
    \includegraphics[width=\linewidth]{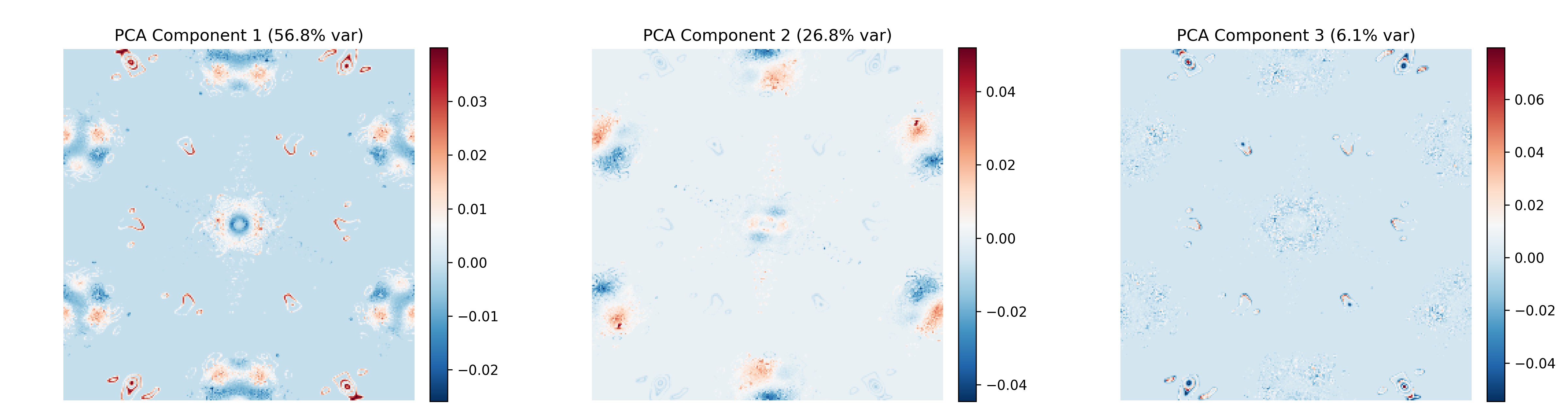}
    \caption{Leading principal components of the FT-LDOS dataset. The top of each panel indicates the fraction of total variance explained by the corresponding component: PC1 (56.8\%), PC2 (26.8\%), and PC3 (6.1\%). Together, the first three components capture 89.7\% of the total variance, demonstrating that the dataset possesses a strongly low-dimensional structure. Color scales represent the signed amplitude of each component in momentum space (arbitrary units).}
    \label{fig:pca_variance}
\end{figure*}

To further analyze the intrinsic structure of the dataset, principal component analysis (PCA) was applied to the flattened FT-LDOS images. The first three principal components capture the majority of the variance, with PC1 accounting for 56.8\%, PC2 for 26.8\%, and PC3 for 6.1\%, corresponding to a cumulative explained variance of 89.7\%. This indicates that despite the high dimensionality of the images, their dominant variations lie on a relatively low-dimensional manifold. The spatial structure of the leading components, shown in Fig.~\ref{fig:pca_variance}, reveals that changes associated with $U$ are concentrated primarily around specific momentum-space regions rather than being uniformly distributed. These observations support the suitability of learning-based approaches, as the relevant information is encoded in a small number of coherent modes rather than in high-dimensional noise.

\begin{figure}[t]
    \centering
    \includegraphics[width=\columnwidth]{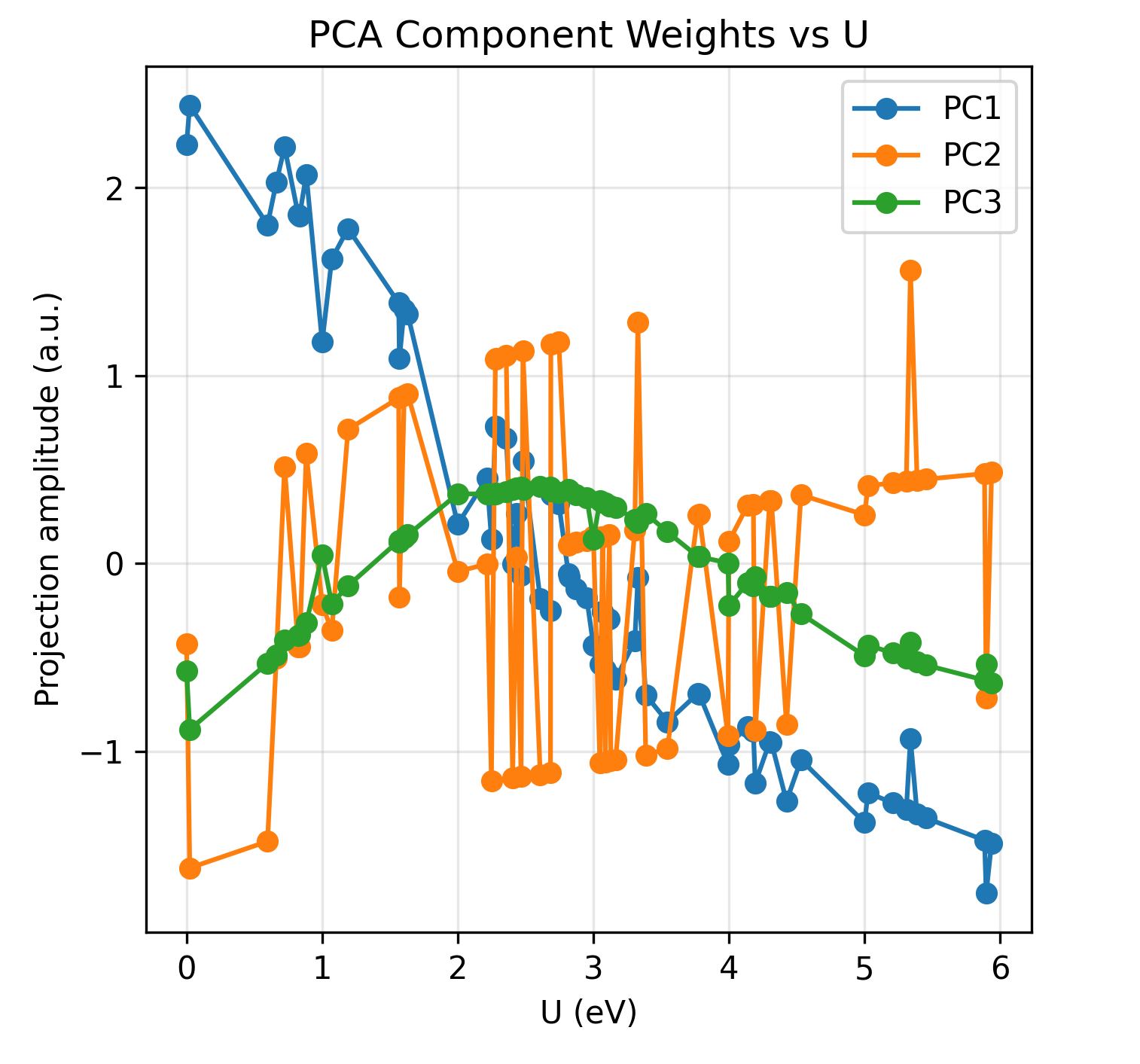}
    \caption{Projection amplitudes of the FT-LDOS images onto the first three principal components as a function of interaction strength $U$. Each point represents the coefficient obtained by projecting the mean image corresponding to a given $U$ value onto the respective PCA eigenvector. The leading component (PC1) exhibits an approximately monotonic dependence on $U$, indicating that a dominant linear feature of the data is directly correlated with the interaction strength. Higher components capture more subtle variations with weaker and less systematic trends.}
    \label{fig:pca_projection}
\end{figure}

The dominance of a small number of principal components suggests that the neural networks primarily need to learn a mapping from a compact latent representation to the interaction parameter $U$, explaining the high regression accuracy obtained despite the limited dataset size. Furthermore, the projection of the FT-LDOS images onto the leading principal component exhibits a nearly monotonic dependence on the interaction strength $U$ (Fig.~\ref{fig:pca_projection}). Inspection of the corresponding eigenvector indicates that PC1 primarily encodes the contrast between the central (moiré) momentum peak and the outer graphene Bragg peaks. This behavior demonstrates that a simple, physically interpretable linear feature already carries substantial information about $U$, providing an intuitive baseline for understanding why the interaction strength can be inferred from STM-derived observables.

\subsection{Regression of Interaction Strength from FT-LDOS}
\label{subsec:regression_int_str}

Model performance was evaluated over three independent random seeds (17, 42, and 101) to assess robustness against data splitting. On the standard test set containing previously seen $U$ values, both architectures achieved consistently high accuracy. The custom CNN obtained an average test performance of MAE=0.14±0.03 eV, RMSE=0.21±0.02 eV, and $R^2$=0.984±0.003. The modified ResNet-18 model yielded comparable results with MAE=0.12±0.02 eV, RMSE=0.21±0.04 eV, and $R^2$=0.985±0.006. These results demonstrate that the mapping from FT-LDOS images to the interaction parameter $U$ can be learned reliably when evaluated within the discrete training manifold.

To place these numbers in context, a linear ridge regression model trained on the leading PCA components attains an average $R^2 \approx
0.91$ on the same test sets. The improved performance of the CNNs, therefore, reflects their ability to exploit nonlinear combinations of momentum-space features beyond the dominant linear trend captured by PCA.

\begin{table}[h]
\caption{Performance metrics averaged over three random seeds (mean ± standard deviation).}
\label{tab:seed_average}
\begin{tabular}{lccc}
\hline
Model & MAE (eV) & RMSE (eV) & $R^2$ \\
\hline
CNN (test) & $0.14 \pm 0.03$ & $0.21 \pm 0.02$ & $0.984 \pm 0.003$ \\
CNN (held-out) & $0.50 \pm 0.08$ & $0.54 \pm 0.07$ & $0.87 \pm 0.04$ \\
ResNet-18 (test) & $0.12 \pm 0.02$ & $0.21 \pm 0.04$ & $0.985 \pm 0.006$ \\
ResNet-18 (held-out) & $0.49 \pm 0.02$ & $0.57 \pm 0.02$ & $0.85 \pm 0.01$ \\
\hline
\end{tabular}
\end{table}

To assess interpolation capability, both models were additionally tested on a held-out dataset containing fractional $U$ values not encountered during training. As expected, performance decreased in this more challenging scenario. The CNN achieved MAE=0.50±0.08 eV and 
$R^2$=0.87±0.04, while ResNet-18 obtained MAE=0.49±0.02 eV and 
$R^2$=0.85±0.01. These results confirm that the models retain meaningful predictive power beyond the discrete grid of training parameters, although with reduced accuracy compared to in-distribution evaluation.

The reduced performance on the held-out set reflects the more challenging task of extrapolating to interaction strengths not encountered during training. This behavior indicates that while the networks successfully interpolate within the training distribution, generalization across previously unseen $U$ values($U > 6$ eV) remains more difficult, resulting in the need for larger and more diverse datasets.

\subsection{CNNs as Detectors of Interaction-Driven Spectral Redistribution}
\label{cnn_detect_inter}

\begin{figure}[t]
    \centering
    \begin{subfigure}[b]{0.32\columnwidth}
        \centering
        \includegraphics[width=\columnwidth]{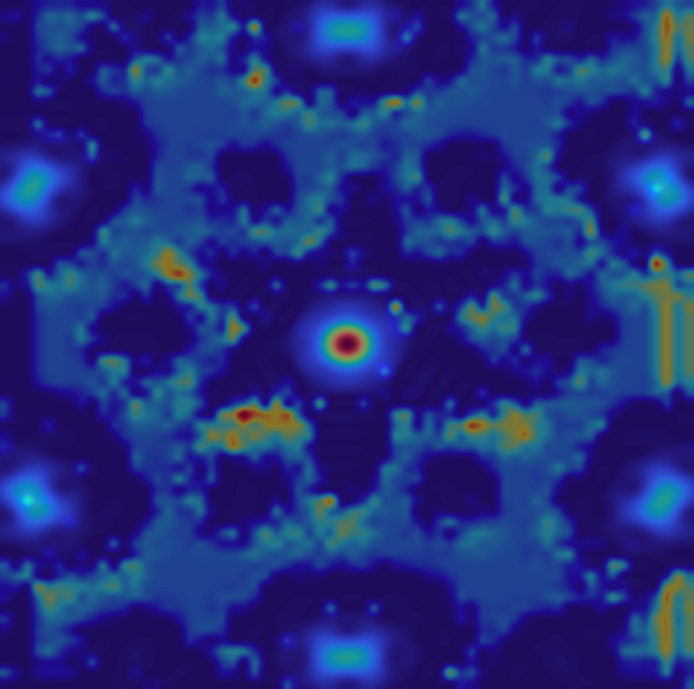}
        \caption{$U=0.0$ eV ($U_{pred}=0.070$ eV)}
        \label{fig:gc_a1}
    \end{subfigure}
    \begin{subfigure}[b]{0.32\columnwidth}
        \centering
        \includegraphics[width=\columnwidth]{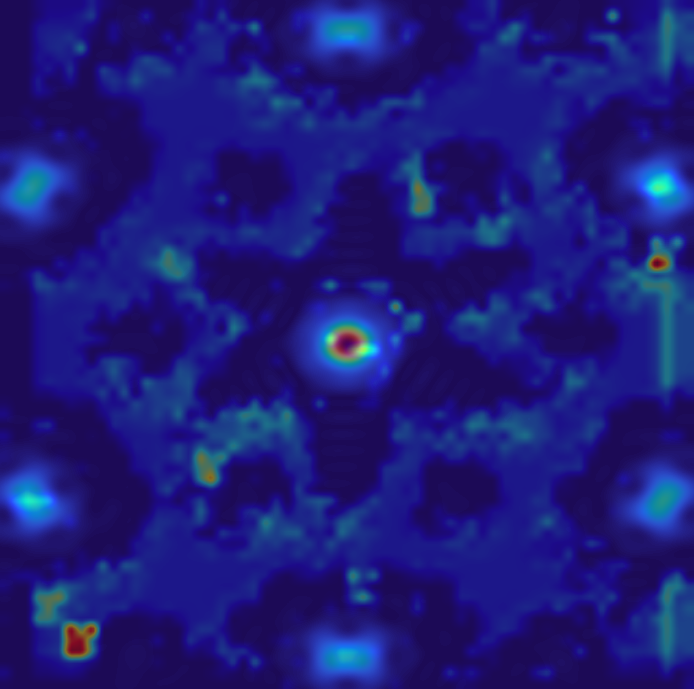}
        \caption{$U=3.0$ eV ($U_{pred}=3.008$ eV)}
        \label{fig:gc_b1}
    \end{subfigure}
    \begin{subfigure}[b]{0.32\columnwidth}
        \centering
        \includegraphics[width=\columnwidth]{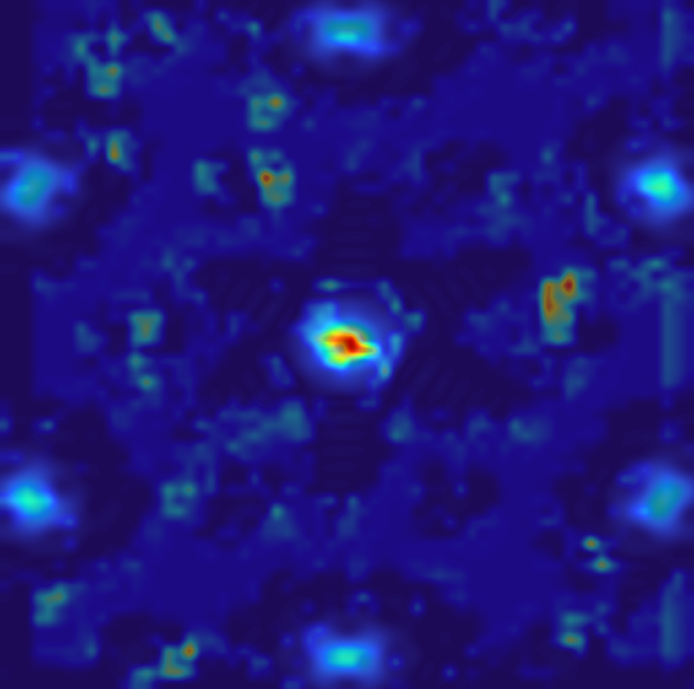}
        \caption{$U=5.0$ eV ($U_{pred}=5.032$ eV)}
        \label{fig:gc_c1}
    \end{subfigure}
    
    \begin{subfigure}[b]{0.32\columnwidth}
        \centering
        \includegraphics[width=\columnwidth]{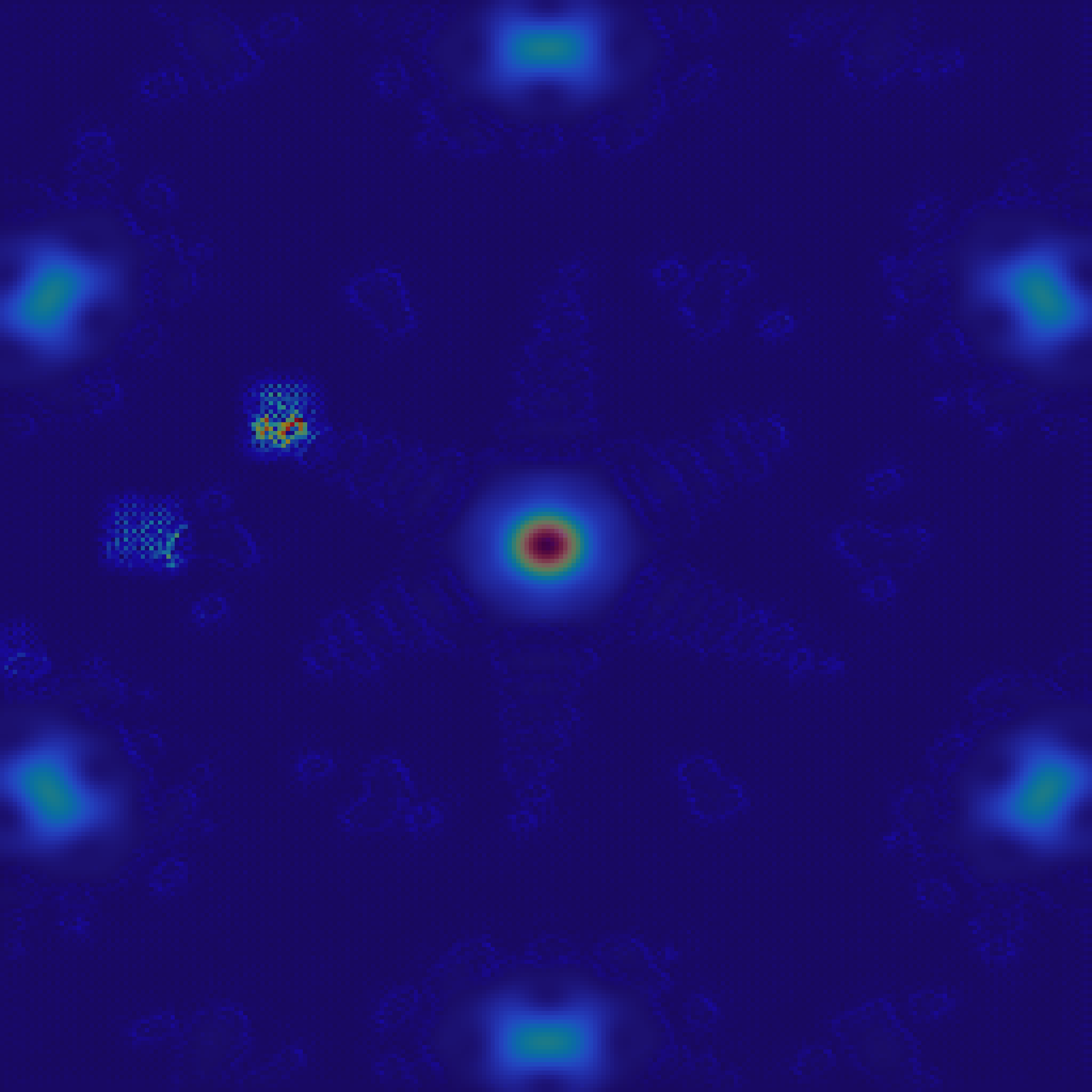}
        \caption{$U=0.0$ eV ($U_{pred}=0.070$ eV)}
        \label{fig:gb_a1}
    \end{subfigure}
    \begin{subfigure}[b]{0.32\columnwidth}
        \centering
        \includegraphics[width=\columnwidth]{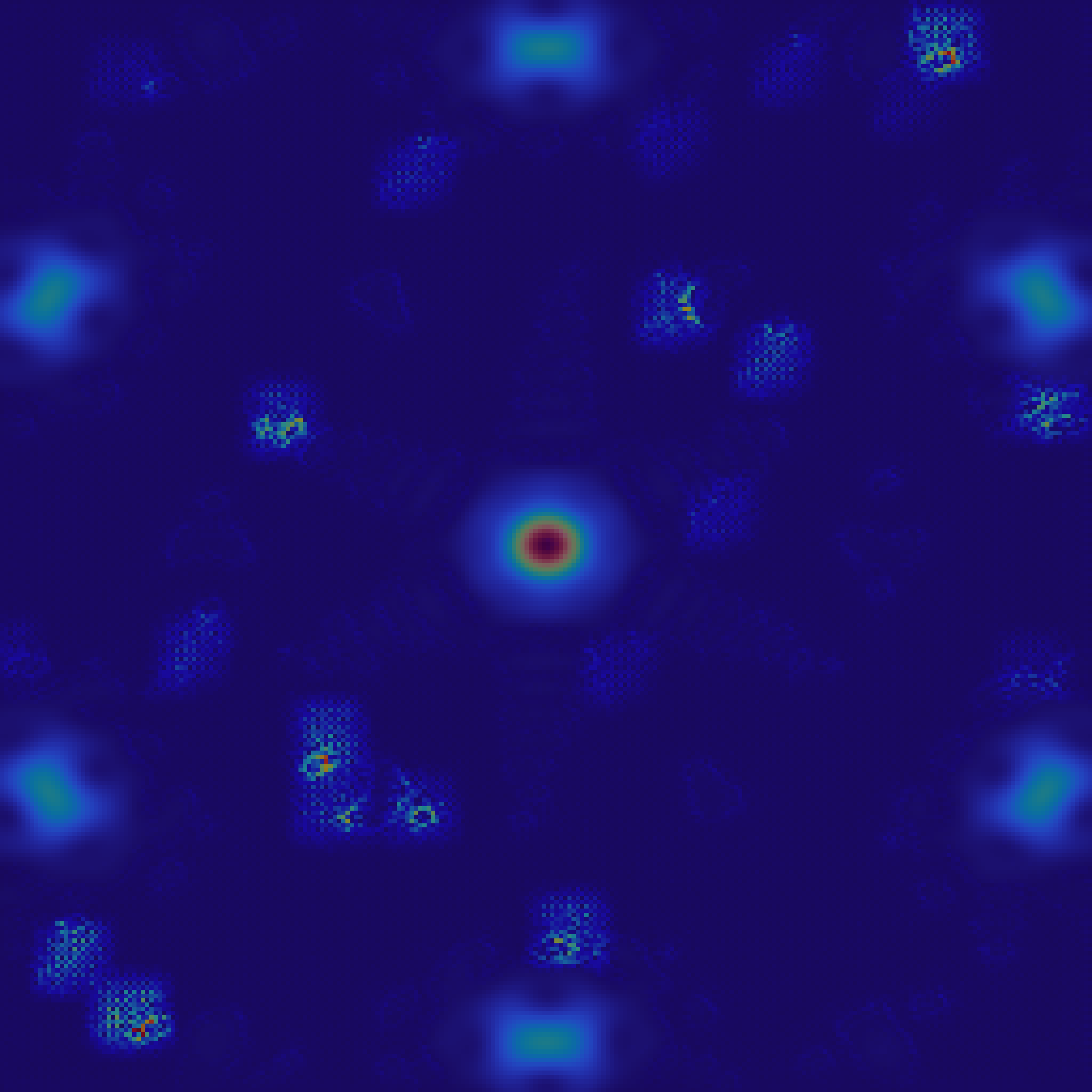}
        \caption{$U=3.0$ eV ($U_{pred}=3.008$ eV)}
        \label{fig:gb_b1}
    \end{subfigure}
    \begin{subfigure}[b]{0.32\columnwidth}
        \centering
        \includegraphics[width=\columnwidth]{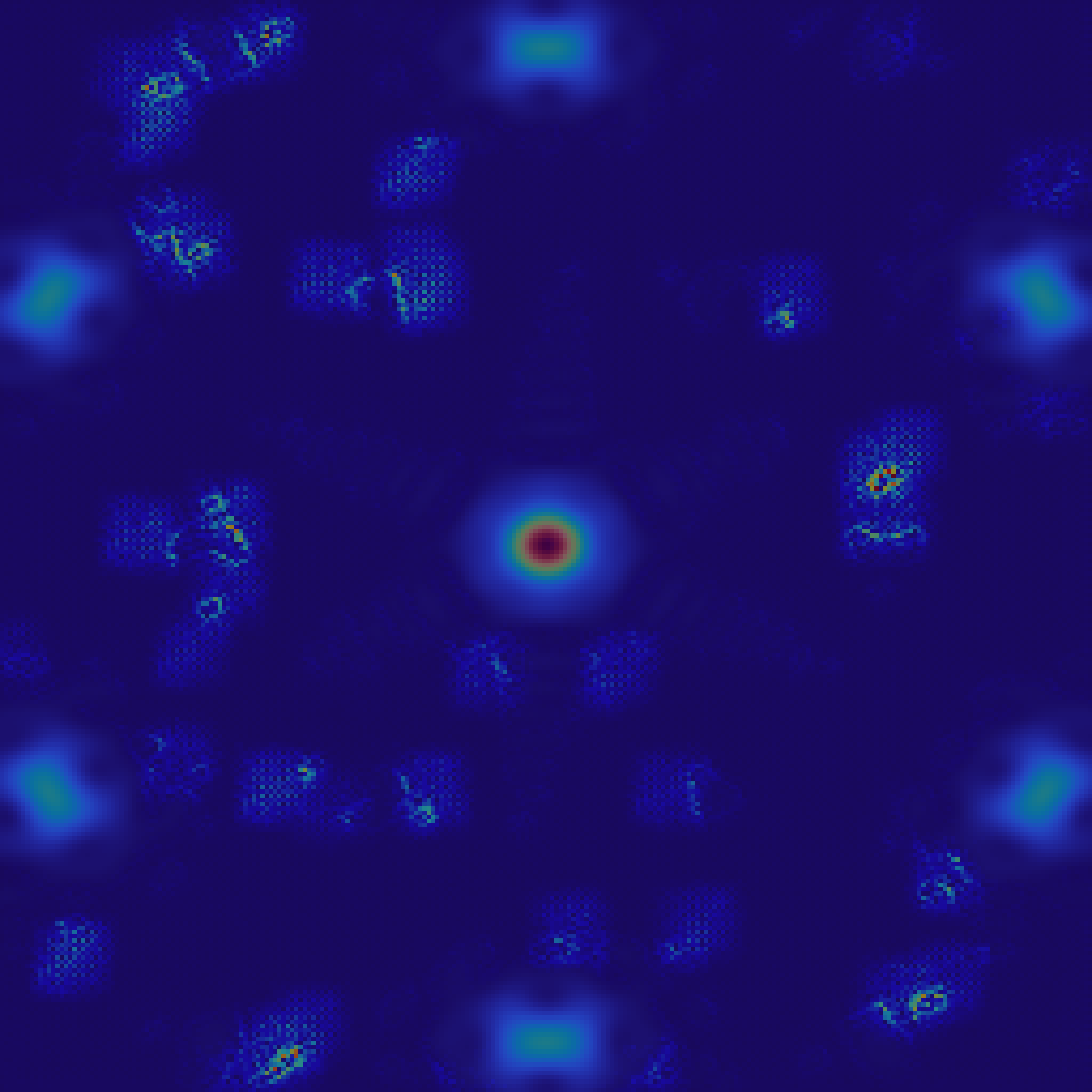}
        \caption{$U=5.0$ eV ($U_{pred}=5.032$ eV)}
        \label{fig:gb_c1}
    \end{subfigure}

    \caption{Momentum-space interpretability of the custom CNN regressor for representative interaction strengths $U$. \textbf{Top row:} Grad-CAM visualizations showing the transition from wide-angle lattice weighting ($U=0.0$) to hyper-localized core focus ($U=5.0$). \textbf{Bottom row:} Guided Backpropagation resolves the pixel-level saliency within the reciprocal lattice peaks. Predicted values ($U_{p}$) are provided in parentheses for each case.}
    \label{fig:interpretability_panel}
\end{figure}

To elucidate which FT-LDOS features control the regression, we apply interpretability tools adapted to regression tasks. Grad-CAM visualizations (Fig.~\ref{fig:interpretability_panel}) reveal a systematic shift in the regions of momentum space most relevant to the predicted interaction strength. At weak coupling ($ U \lesssim 1$ eV), the CNNs attend broadly to both central and satellite moiré peaks. As the interaction increases to intermediate values ($ U \approx 2-3$ eV), attention becomes increasingly focused on the relative intensity contrast between these features. In the strong-coupling regime (
$ U \gtrsim 4$ eV), the response localizes predominantly near the $\Gamma$-point region.

Guided backpropagation further resolves these trends at the pixel level. Saliency maps indicate that the CNNs respond primarily to sharp gradients at the edges of reciprocal lattice peaks rather than to diffuse background regions, confirming sensitivity to interaction-induced changes in peak width and relative spectral weight. Regions of stochastic background contribute negligibly to the regression signal.

\subsection{Robustness Against Experimental Perturbations}

A potential limitation of training exclusively on idealized theoretical data is that the machine-learning model may learn features specific to the simulation pipeline rather than physically robust characteristics of the FT-LDOS. To investigate this possibility, we evaluated both linear
regression and convolutional neural-network models under controlled perturbations that emulate experimental STM imperfections.

The models were trained and tested on combinations of clean, mildly perturbed, symmetry-breaking, and noisy datasets. Table~\ref{tab:robustness_summary} summarizes the most relevant results.

For both the standard test set and the holdout set, models trained exclusively on clean data exhibit a substantial performance degradation when evaluated on perturbed images. In contrast, training with physically motivated augmentations significantly improves robustness while maintaining high accuracy on the original clean data.

The custom convolutional neural network demonstrates the best overall trade-off between accuracy and robustness. In particular, training on the mildly perturbed dataset yields strong performance on clean, mildly perturbed, and symmetry-breaking images, achieving $R^2=0.972$, $R^2=0.950$, and $R^2=0.866$, respectively, on the independent test set. Similar trends are observed on the holdout dataset containing previously unseen interaction strengths.

These results indicate that the CNN does not merely learn to invert the idealized forward model. Instead, the network learns physically meaningful momentum-space features that remain informative under moderate experimental distortions. A complete ablation study for the linear baseline and custom CNN is provided in Appendix~\ref{app:robustness}.

\begin{table}[t]
\caption{
Summary of robustness results for the custom CNN. Complete ablation results are reported in Appendix~\ref{app:robustness}.
}
\label{tab:robustness_summary}
\centering
\begin{tabular}{lcc}
\hline
Train $\rightarrow$ Test & MAE & $R^2$ \\
\hline
Clean $\rightarrow$ Clean & 0.11 & 0.988 \\
Clean $\rightarrow$ Mild & 3.04 & -4.39 \\
Mild $\rightarrow$ Clean & 0.25 & 0.972 \\
Mild $\rightarrow$ Mild & 0.31 & 0.950 \\
Mild $\rightarrow$ Symmetry-breaking & 0.53 & 0.866 \\
Symmetry-breaking $\rightarrow$ Symmetry-breaking & 0.24 & 0.935 \\
\hline
\end{tabular}
\end{table}

\subsection{Indications of an Interaction-Dependent Spectral Crossover}
As already mentioned, the interpretability and PCA analyses suggest a gradual interaction-dependent redistribution of spectral weight, with a noticeable change in behavior around $U \sim t=2.7$eV. Also, from the principal component analysis, we can draw this conclusion. In Fig. \ref{fig:pca_projection}, one sees that PC1 as a function of $U$ is monotonically decreasing with a zero at $U\sim t=2.7$eV. PC2 is more fluctuating and can be decomposed into three main branches, all monotonically increasing, and the middle one with a zero again at $U\sim t=2.7$eV. Finally, PC3 displays a maximum at  $U\sim t=2.7$eV.

The atomistic Hubbard interaction has important implications for the emergence of magnetic phases in correlated systems. For single-layer graphene at half filling, antiferromagnetism is stable beyond a critical Hubbard interaction $U_c/t\sim2.23$.\cite{Sorella92,Assaad13} For twisted bilayer graphene at the magic angle, this critical value is considerably reduced and estimated to be $U_c/t \sim0.23$,\cite{Vahedi21} consistent with the approach of Ref.~\onlinecite{Gonzalez20b} and still sufficient to induce ferromagnetism.\cite{Pons20} Our analysis suggests that even for unpolarized electronic states, these observations suggest a possible crossover scale of order $U/t \sim 1$, although identifying a true critical interaction would require a dedicated analysis of symmetry-broken Hartree--Fock solutions and is beyond the scope of the present work.

\section{Summary}
We have implemented a CNN for image recognition of STM data. Our results demonstrate that FT-LDOS images obtained from realistic Hartree--Fock modeling can be used to quantitatively infer effective interaction strengths in correlated moiré materials.
The above discussion also shows that the CNNs are sensitive to interaction-dependent spectral redistribution, achieving high prediction accuracy (with coefficients of determination $R^2 \gtrsim 0.95$ on test data) and robust interpolation for fractional $U$ values near the critical regime, where the dominant FT-LDOS weight shifts from central moiré peaks to satellite ordering vectors.\cite{Sobral_2023,Taranto22}

Furthermore, we investigated the robustness of the machine-learning models against physically motivated perturbations designed to emulate realistic STM imperfections. While models trained exclusively on idealized FT-LDOS images exhibit a substantial performance degradation under distribution shifts, augmentation-based training significantly improves robustness to both noisy and symmetry-breaking inputs. These results suggest that the learned representation is not solely tied to the exact symmetries of the simulated dataset and may remain applicable under realistic
experimental conditions. This establishes a practical route to inferring effective interaction strengths directly from STM data in correlated moiré materials, and paves the way for combined theory ML-STM frameworks to quantify microscopic parameters in future experiments on twisted multilayers and related systems.

We believe that methodology is readily transferable to experimental STM data on twisted bilayers and multilayers, and can be extended to other parameters such as screening length, strain, or twist-angle inhomogeneity, providing a powerful, data-driven route to Hamiltonian learning in quantum materials.

The present study remains limited by the use of synthetic Hartree--Fock-generated training data. Although augmentation-based training improves robustness against moderate perturbations and symmetry-breaking distortions, further work will be required to evaluate transferability to fully experimental STM datasets containing uncontrolled instrumental and sample-dependent effects.

\section*{Data Availability}
All code used for dataset generation, model training, evaluation, and analysis is openly available in Zenodo at \url{https://doi.org/10.5281/zenodo.18717017}. The source code is also maintained on GitHub at \url{https://github.com/nachiket273/tbg-stm-ftldos-hubbard-U-regression}.

{\it Acknowledgements.}
The work was supported by a grant PID2023-146461NB-I00 funded by MCIN/AEI/10.13039/501100011033 as well as by the CSIC Research Platform on Quantum Technologies PTI-001. The access to computational resources of CESGA (Centro de Supercomputación de Galicia) is also gratefully acknowledged.

%

\appendix

\section{Training Hyper-Parameters}
\label{app:hyperparameters}

This appendix summarizes the hyperparameter values used for all neural network models reported in the main text. The values listed here were held constant across all training runs and random data splits unless otherwise stated.

All convolutional neural networks were trained using the Adam optimizer. The initial learning rate was set to $10^{-4}$ for the custom CNN and to $10^{-3}$ for the modified ResNet-18 architecture. The optimizer momentum parameters were $\beta_1 = 0.9$ and $\beta_2 = 0.999$. A cosine annealing learning-rate schedule was employed, with a minimum learning rate of $10^{-6}$. The batch size was fixed at 32 for all experiments.

Models were trained for up to 40 epochs, with early stopping based on validation loss, using a patience of 10 epochs. A weight decay of $10^{-4}$ was applied to mitigate overfitting. For the baseline clean-data models, no data augmentation was used. For the robustness analysis in Appendix~B, additional models were trained using the augmentation protocols described in \ref{subsec: robust_eval}. To ensure numerical stability during training, gradient clipping with a maximum norm of 1.0 was applied.

The mean squared error between the predicted and target interaction strengths was used as the training loss. Model performance was evaluated using the mean absolute error (MAE), root-mean-square error (RMSE), and the coefficient of determination $R^2$, as reported in Table~\ref{tab:seed_average}. All reported results correspond to the same hyper-parameter configuration across different random training-test splits.

Training was performed using fixed random seeds (17, 42, and 101) to ensure reproducibility. The full implementation, including configuration files specifying all hyperparameters and random seeds, will be made publicly available upon publication.

\section{Robustness Ablation Study}
\label{app:robustness}

\subsection{Augmentation Datasets}

To evaluate the robustness of the machine-learning models against experimental imperfections, three additional datasets were generated from the original FT-LDOS images.

\begin{itemize}
\item \textbf{Mild perturbation dataset}: weak Gaussian noise, gain variations, anisotropic blurring, and scan-line artifacts designed to emulate realistic STM measurements while preserving the dominant FT-LDOS features.

\item \textbf{Symmetry-breaking dataset}: small affine distortions introduced to mimic weak heterostrain, calibration errors, or geometric distortions that reduce the ideal $C_3$ symmetry of the simulated images.

\item \textbf{Noisy dataset}: a stronger perturbation regime combining scan-line noise, anisotropic blur, gain variations, and stochastic noise. This dataset serves as a stress test for model generalization under significant distribution shifts.
\end{itemize}

Representative examples are shown in Fig.~\ref{fig:augmentation_examples}.

\subsection{Robustness Analysis on Test Set}

\begin{table*}[t]
\caption{
Linear regression robustness study on the standard test set.
}
\label{tab:linear_test}
\centering
\begin{tabular}{lcc}
\hline
Train $\rightarrow$ Test & MAE & $R^2$ \\
\hline
Clean $\rightarrow$ Clean & 0.034 & 0.998 \\
Clean $\rightarrow$ Mild & 0.343 & 0.771 \\
Clean $\rightarrow$ Symmetry-breaking & 10.334 & -42.831 \\
Clean $\rightarrow$ Noisy & 1.158 & -0.162 \\
\hline
Mild $\rightarrow$ Clean & 0.052 & 0.995 \\
Mild $\rightarrow$ Mild & 0.344 & 0.852 \\
Mild $\rightarrow$ Symmetry-breaking & 6.290 & -15.299 \\
Mild $\rightarrow$ Noisy & 1.007 & 0.249 \\
\hline
Symmetry-breaking $\rightarrow$ Clean & 0.152 & 0.984 \\
Symmetry-breaking $\rightarrow$ Mild & 0.489 & 0.662 \\
Symmetry-breaking $\rightarrow$ Symmetry-breaking & 0.213 & 0.950 \\
Symmetry-breaking $\rightarrow$ Noisy & 1.329 & -0.606 \\
\hline
Noisy $\rightarrow$ Clean & 0.091 & 0.984 \\
Noisy $\rightarrow$ Mild & 0.320 & 0.874 \\
Noisy $\rightarrow$ Symmetry-breaking & 47.706 & -969.062 \\
Noisy $\rightarrow$ Noisy & 0.654 & 0.625 \\
\hline
\end{tabular}
\end{table*}

Table~\ref{tab:linear_test} summarizes the test set performance of the linear regression under different training and testing conditions. The linear regression model exhibits excellent performance when the training and testing distributions match. However, substantial degradation is observed under distribution shifts, particularly for symmetry-breaking and strongly perturbed images. This indicates that linear regression relies heavily on the exact structure of the simulated FT-LDOS images and has limited robustness to experimental distortions.

\begin{table*}[t]
\caption{
Robustness study of the custom CNN on the standard test set.
}
\label{tab:cnn_test}
\centering
\begin{tabular}{lcc}
\hline
Train $\rightarrow$ Test & MAE & $R^2$ \\
\hline
Clean $\rightarrow$ Clean & 0.111 & 0.988 \\
Clean $\rightarrow$ Mild & 3.042 & -4.387 \\
Clean $\rightarrow$ Symmetry-breaking & 2.645 & -2.935 \\
Clean $\rightarrow$ Noisy & 1.085 & 0.157 \\
\hline
Mild $\rightarrow$ Clean & 0.255 & 0.972 \\
Mild $\rightarrow$ Mild & 0.311 & 0.950 \\
Mild $\rightarrow$ Symmetry-breaking & 0.534 & 0.867 \\
Mild $\rightarrow$ Noisy & 0.531 & 0.811 \\
\hline
Symmetry-breaking $\rightarrow$ Clean & 0.243 & 0.936 \\
Symmetry-breaking $\rightarrow$ Mild & 1.086 & -0.116 \\
Symmetry-breaking $\rightarrow$ Symmetry-breaking & 0.237 & 0.935 \\
Symmetry-breaking $\rightarrow$ Noisy & 1.579 & -0.744 \\
\hline
Noisy $\rightarrow$ Clean & 0.258 & 0.949 \\
Noisy $\rightarrow$ Mild & 0.389 & 0.894 \\
Noisy $\rightarrow$ Symmetry-breaking & 0.304 & 0.943\\
Noisy $\rightarrow$ Noisy & 0.517 & 0.819 \\
\hline
\end{tabular}
\end{table*}

Table~\ref{tab:cnn_test} summarizes the test set performance of the custom CNN under different training and testing conditions. Compared with linear regression, the CNN exhibits significantly improved robustness under moderate distribution shifts. In particular, training on the mildly perturbed dataset results in strong performance on clean, perturbed, and symmetry-breaking images simultaneously. The model achieves $R^2 = 0.972$ on clean images, $R^2 = 0.950$ on mildly perturbed images, and $R^2 = 0.867$ on symmetry-breaking images, demonstrating that the learned representation is not strictly tied to the exact symmetry of the simulated training data.

\subsection{Holdout Set Analysis}

The holdout dataset provides a more stringent evaluation of model generalization because it contains interaction strengths that were not present during training. Unlike the standard test set, which samples the same discrete interaction values used for training, the holdout set probes the ability of the model to interpolate between previously observed interaction strengths.

Table~\ref{tab:holdout_linear} summarizes the holdout-set performance of the linear regression under different training and testing conditions. The linear regression model maintains reasonable performance when the train and test distributions are similar, achieving $R^2 \approx 0.87$ on the clean holdout dataset. However, substantial performance degradation is observed under symmetry-breaking and strongly perturbed conditions, indicating a limited ability to generalize under simultaneous parameter and distribution shifts.

\begin{table*}[t]
\caption{
Linear regression robustness analysis on the holdout dataset.
}
\label{tab:holdout_linear}
\centering
\begin{tabular}{lcc}
\hline
Train $\rightarrow$ Test & MAE & $R^2$ \\
\hline
Clean $\rightarrow$ Clean & 0.449 & 0.873 \\
Clean $\rightarrow$ Mild & 0.585 & 0.654 \\
Clean $\rightarrow$ Noisy & 1.550 & -1.902 \\
Clean $\rightarrow$ Symmetry-breaking & 8.796 & -48.145 \\
\hline
Mild $\rightarrow$ Clean & 0.466 & 0.869 \\
Mild $\rightarrow$ Mild & 0.515 & 0.791 \\
Mild $\rightarrow$ Noisy & 1.292 & -0.617 \\
Mild $\rightarrow$ Symmetry-breaking & 5.220 & -16.117 \\
\hline
\end{tabular}
\end{table*}

\begin{table*}[!t]
\caption{
CNN robustness analysis on the holdout dataset.
}
\label{tab:holdout_cnn}
\centering
\begin{tabular}{lcc}
\hline
Train $\rightarrow$ Test & MAE & $R^2$ \\
\hline
Clean $\rightarrow$ Clean & 0.5085 & 0.8601 \\
Clean $\rightarrow$ Mild & 2.8452 & -4.6835 \\
Clean $\rightarrow$ Symmetry-breaking & 3.3343 & -5.5440 \\
Clean $\rightarrow$ Noisy & 3.5035 & -6.2208 \\
\hline
Mild $\rightarrow$ Clean & 0.4078 & 0.9080 \\
Mild $\rightarrow$ Mild & 0.4358 & 0.8825 \\
Mild $\rightarrow$ Symmetry-breaking & 0.3252 & 0.9070 \\
Mild $\rightarrow$ Noisy & 0.6312 & 0.5764 \\
\hline
\end{tabular}
\end{table*}

Table~\ref{tab:holdout_cnn} summarizes the holdout-set performance of the custom CNN under different training and testing conditions. The CNN exhibits similar qualitative trends. Models trained exclusively on idealized FT-LDOS images show a significant reduction in performance when applied to perturbed holdout images. In contrast, augmentation-based training improves generalization across both the interaction-strength and perturbation domains. The mildly perturbed training set consistently provides the best compromise between interpolation accuracy and robustness to realistic experimental distortions.

These results suggest that the learned representation is not solely dependent on the discrete interaction values present in the training set. Instead, the CNN captures continuous interaction-dependent features of the FT-LDOS that remain informative for previously unseen interaction strengths and moderate distribution shifts.

\subsection{Summary of Robustness Results}

Across all experiments, the strongest generalization performance is obtained when the CNN is trained using the mildly perturbed dataset. This training strategy preserves high accuracy on clean
FT-LDOS images while simultaneously improving robustness to symmetry-breaking distortions and
moderate experimental perturbations. These observations suggest that augmentation-based
training reduces dependence on idealized symmetries and encourages the extraction of physically meaningful momentum-space features.

The robustness trends observed for both the standard test set and the holdout dataset provide additional evidence that the CNN is not merely memorizing the forward simulation pipeline but is instead learning transferable representations associated with the interaction-dependent structure of the FT-LDOS.

\appendix
\end{document}